\documentclass[aps,prl,twocolumn,amsmath,amssymb, superscriptaddress]{revtex4-2}
\usepackage{graphicx}
\graphicspath{{Figures/}}
\usepackage{subfigure}
\usepackage{epsfig}
\usepackage{ulem}
\usepackage{dcolumn}
\usepackage{bm}
\usepackage{hyperref}
\hypersetup{colorlinks=true, citecolor=blue, urlcolor=blue, linkcolor=blue}
\usepackage{appendix}

\renewcommand{\v}[1]{{\boldsymbol{#1}}}

\newcommand{\imth}{\hspace{1pt}\mathrm{i}\hspace{1pt}}
\newcommand{\dif}{\mathrm{d}}
\newcommand{\Dif}{\mathrm{D}}

\newcommand{\E}[1]{{\mathrm{e}^{ #1}}}

\newcommand{\Tr}{{\rm Tr}}
\newcommand{\Det}{{\rm Det}}

\newcommand{\ra}{\rightarrow}
\newcommand{\s}{{\sigma}}

\def\beq{\begin{equation}}
\def\eeq{\end{equation}}
\def\bald{\begin{aligned}}
\def\eald{\end{aligned}}
\def\bea{\begin{eqnarray}}
\def\eea{\end{eqnarray}}

\def\inc#1{\left(#1\right)}
\def\Inc#1{\left[#1\right]}

\def\avg#1{\left\langle#1\right\rangle}

\def\Tr{\mathrm{Tr}}

\def\Eq#1{Eq.~(\ref{#1})}
\def\Fig#1{Fig.~\ref{#1}}

\begin{document}
\title{High-temperature superconductivity induced by the Su-Schrieffer-Heeger electron-phonon coupling}
\author{Xun Cai}
\affiliation{Beijing National Laboratory for Condensed Matter Physics and Institute of Physics, Chinese Academy of Sciences, Beijing 100190, China}
\author{Zi-Xiang Li}
\email{zixiangli@iphy.ac.cn}
\affiliation{Beijing National Laboratory for Condensed Matter Physics and Institute of Physics, Chinese Academy of Sciences, Beijing 100190, China}
\affiliation{University of Chinese Academy of Sciences, Beijing 100049, China}
\author{Hong Yao}
\email{yaohong@tsinghua.edu.cn}
\affiliation{Institute for Advanced Study, Tsinghua University, Beijing 100084, China}

\begin{abstract}
Experimental quest for high-temperature and room-temperature superconductivity (SC) at ambient pressure has been a long-standing research theme in physics.
It has also been desired to construct reliable microscopic mechanisms that may achieve high-temperature SC.
Here we systematically explore SC in the Su-Schrieffer-Heeger (SSH) electron-phonon coupling models by performing numerically-exact quantum Monte-Carlo simulations. 
Our results reliably showed that superconducting $T_c$ of the SSH models is high, remarkably higher than those in the Holstein models, particularly in strong electron-phonon coupling regime.
This is mainly because SSH phonons can not only induce strong pairing between electrons but also help the phase coherence of Cooper pairs, thus realizing higher $T_c$.
As mechanism of higher-$T_c$ of the SSH models could be potentially relevant to realistic materials, it paves a promising way to find higher-temperature SC in the future.
\end{abstract}
\date{\today}

\maketitle
\textbf{Introduction}:
Pursuit for superconductivity (SC) with increasingly higher transition temperature at ambient pressure has been one of the central topics in physics \cite{KeimerReview,Kivelson2003RMP,ReviewSachdev2003,Wen2006RMP,ironbasedReview,DHLee2018Review,Ding2022Review}. 
In the BCS theory \cite{BCSoriginal, Schrieffer1964book}, electron-phonon coupling (EPC) is crucial for Cooper pairing in conventional superconductors. However, EPC in many conventional superconductors is relatively weak, giving rise to low transition temperature ($T_c$).
Even when EPC is strong, high temperature SC is usually hindered by various constraints \cite{Kivelson2018npjQM}, such as competing electronic/lattice instabilities and formation of heavy bipolarons \cite{Scalapino1989PRB,Cohen2006PRB, Alexandrov2001EPLbipolaron,Prokofev1998PRLPolaron, Bonca1999PRBPolaron,Chakraverty1998PRLPolaron}.
For instance, for the Holstein electron-phonon model
\cite{Holsteinoriginal,Scalettar1991PRLHolstein, Scalapino1993PRBHolstein,Sawatzky1994Holstein, Berger1995PRBHolstein,White1998PRBHolstein, Marsiglio1990PRBHolstein, Clay2005PRLHolstein,Millis2007EPC,Johnston2013PRBHolstein, ZXLi2019PRBHolstein, Wang2015Holstein,Kivelson2018PRBHolstein,Kivelson2019PRBHolstein, Scalettar2018PRLHolstein,Meng2019PRLHolstein, Batrouni2020PRBHolstein, ZYHan2020PRL,Devereaux2021PRBHolstein,ZXLi2019PRBFeSe,Wang2020EPC,Scalettar2018PRL},  recent quantum Monte-Carlo (QMC) simulations showed that competing charge-density wave (CDW) instabilities emerges with the formation of heavy bipolarons and strongly suppressed $T_c$ \cite{Kivelson2018PRBHolstein,Kivelson2019PRBHolstein} for strong Holstein EPC. Realizing high-$T_c$ SC at ambient pressure dominantly driven by EPC remains a challenge, although intriguing progresses towards this goal have been made \cite{XueFeSe,MgB2Nature,DHLee2007PRB-fRG-dwave,Xiang2015PRB,Pickett2001PRL,Feng2019NC,Shen2015Nature,Li2016SB,Shen2021Science,Tang2023NC,Wang2022arXivcuprate,Li2019PRL}.

\begin{figure}[t]
\includegraphics[width= 0.455\linewidth]{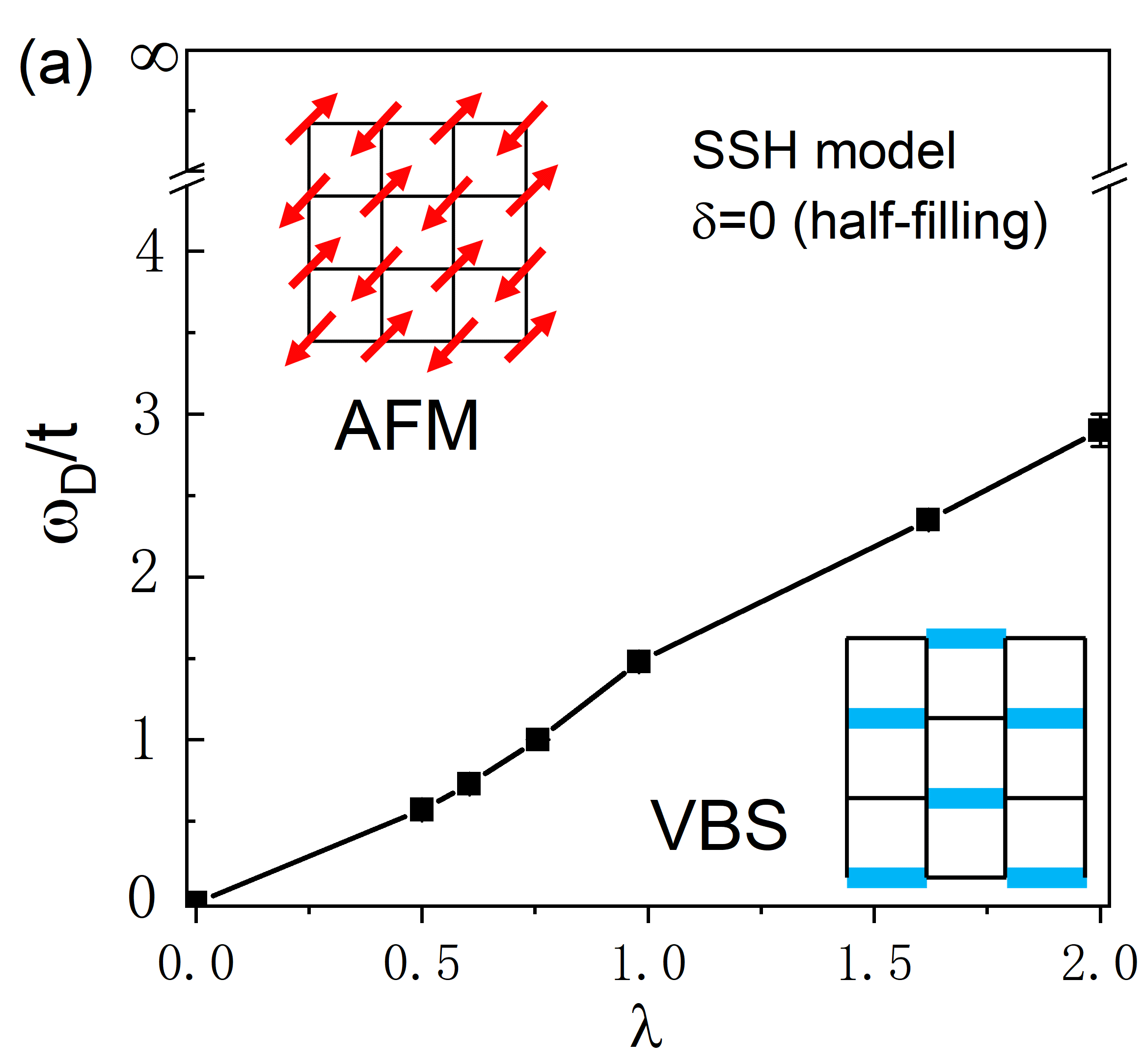}~
\includegraphics[width= 0.475\linewidth]{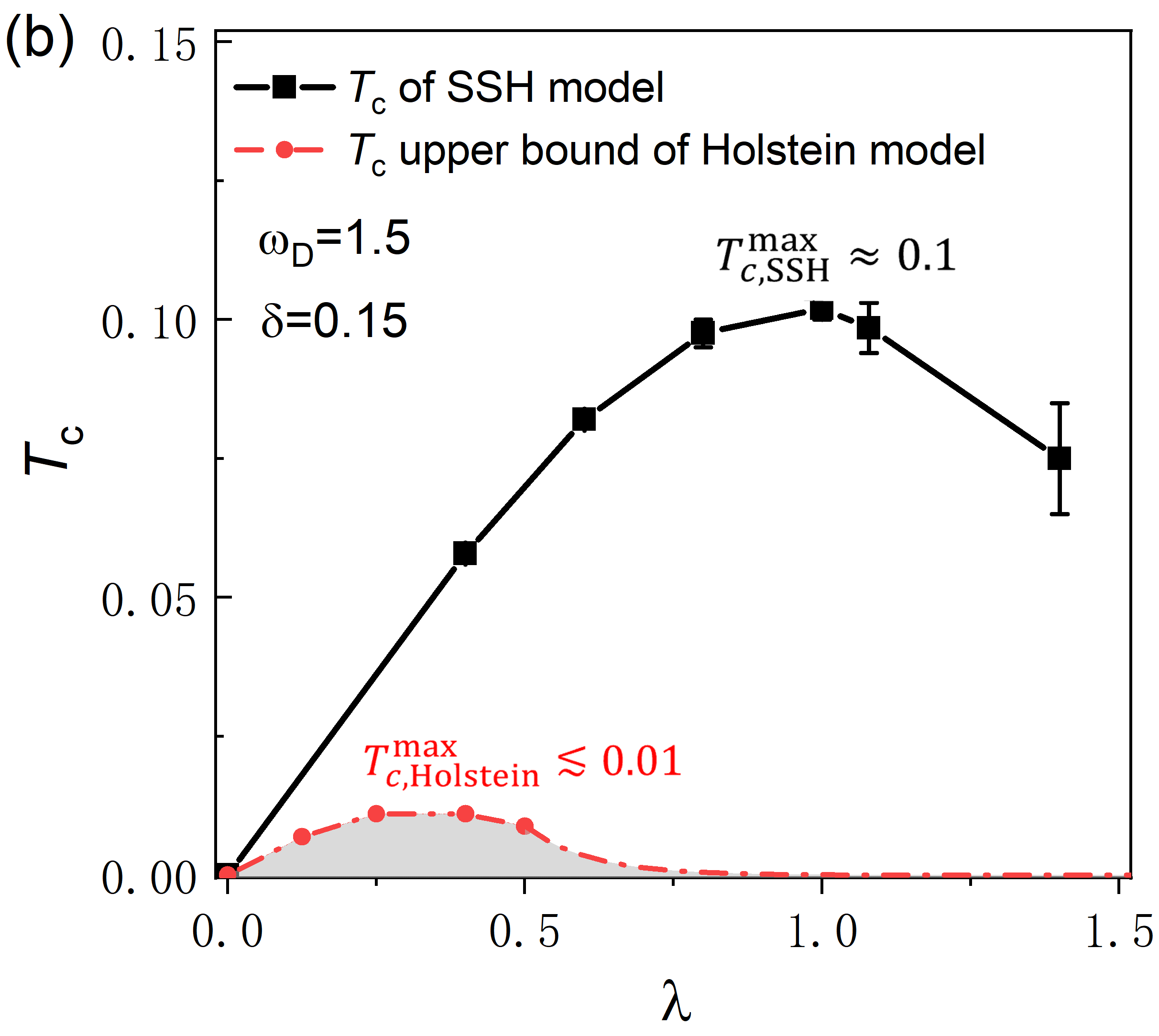}
\caption{(a) Quantum phase diagram of the SSH model at half-filling obtained by three of us in Ref. \cite{XC2021PRL}. Here $\omega_D$ and $\lambda$ are phonon frequency and EPC strength, respectively. (b) Superconducting $T_c$ as the function of EPC strength $\lambda$ in SSH model for fixed phonon frequency $\omega_D=1.5$ and hole doping concentration $\delta=0.15$. The red line refers to the upper bound of $T_c$ of the Holstein model obtained from QMC simulations (all
energies are in units of $t=1$).}
\label{Fig1}
\end{figure}

In contrast with Holstein phonons which couple with electron density, Su-Schrieffer-Heeger (SSH) phonons couple with electron hopping \cite{originalSSH, SSHRMP}.
Recently, SSH models have received increasing interest \cite{Sous2018, Sous2021, Assaad2015SSH, Hohenadler2020SSH, Scalettar2021PRL, XC2021PRL, Cai2022PRBSSH, Assaad2022PRBSSH, Scalettar2022PRB, Assaad2023arXivSSH,Johnston2019npj,Johnston2022arXiv,Wang2022PRB,Wang2022arXiv1,Sous2023arXiv,Scalettar2023SSHHarXiv,Johnston2023compareSSHarXiv,Johnston2023SCarXiv} partly because they have various intriguing properties compared with Holstein models.
For instance, the SSH model on square lattice at half filling can induce anti-ferromagnetism (AFM) or CDW/SC in ground state for phonon frequency above a critical value \cite{XC2021PRL,Cai2022PRBSSH,Assaad2022PRBSSH, Scalettar2022PRB,Assaad2023arXivSSH}, as shown in \Fig{Fig1}(a). 
Moreover, recent studies showed that bipolarons in the 1D SSH model can be significantly lighter than those in the Holstein model, implying a possible route to higher temperature SC from SSH phonons \cite{Sous2018}; evidences of such light bipolarons were obtained by QMC simulation in dilute limit of electron filling \cite{Millis2023PRX,Zhang2022PRBQMC,Zhang2021PRB}.
However, it remains to show whether superconductivity with high-$T_c$ can emerge in the 2D SSH model at generic filling from numerically-exact approaches such as QMC.

In this Letter, we perform large-scale QMC simulations of the square lattice SSH model at finite doping to explore its possible high-$T_c$ superconductivity.
As this model at any doping is free from the notorious sign problem \cite{CJWu2005PRB,ZXLi2015PRB, ZXLi2016PRL, TXiang2016PRL, LWang2015PRL, ZXLiQMCreview}, we can reliably simulate it with sufficiently large lattice size and low temperature to accurately obtain its $T_c$. 
For comparison, we also studied $T_c$ of the Holstein model. Remarkably, we find that the superconducting $T_c$ of the SSH model is substantially higher than that of the Holstein model for the same phonon frequency $\omega_D$ and EPC  $\lambda$, especially in the strong coupling regime, as shown in \Fig{Fig1}(b). Moreover, for a fixed $\omega_D$, $T_c$ of the SSH model exhibits a dome-like behavior as a function of $\lambda$; the maximum $T_{c,\textrm{SSH}}^{\textrm{max}}$ occurs at $\lambda_o$ that is close to the critical value $\lambda_c$ separating the VBS and AFM at half filling. For $\omega_D=1.5$ and hole doping $\delta=0.15$, $T_{c,\textrm{SSH}}^{\textrm{max}}\approx 0.1$, a huge increase from $T_{c,\textrm{Holstein}}^{\textrm{max}} < 0.01$.

\begin{figure}[t]
\includegraphics[width= 0.46\linewidth]{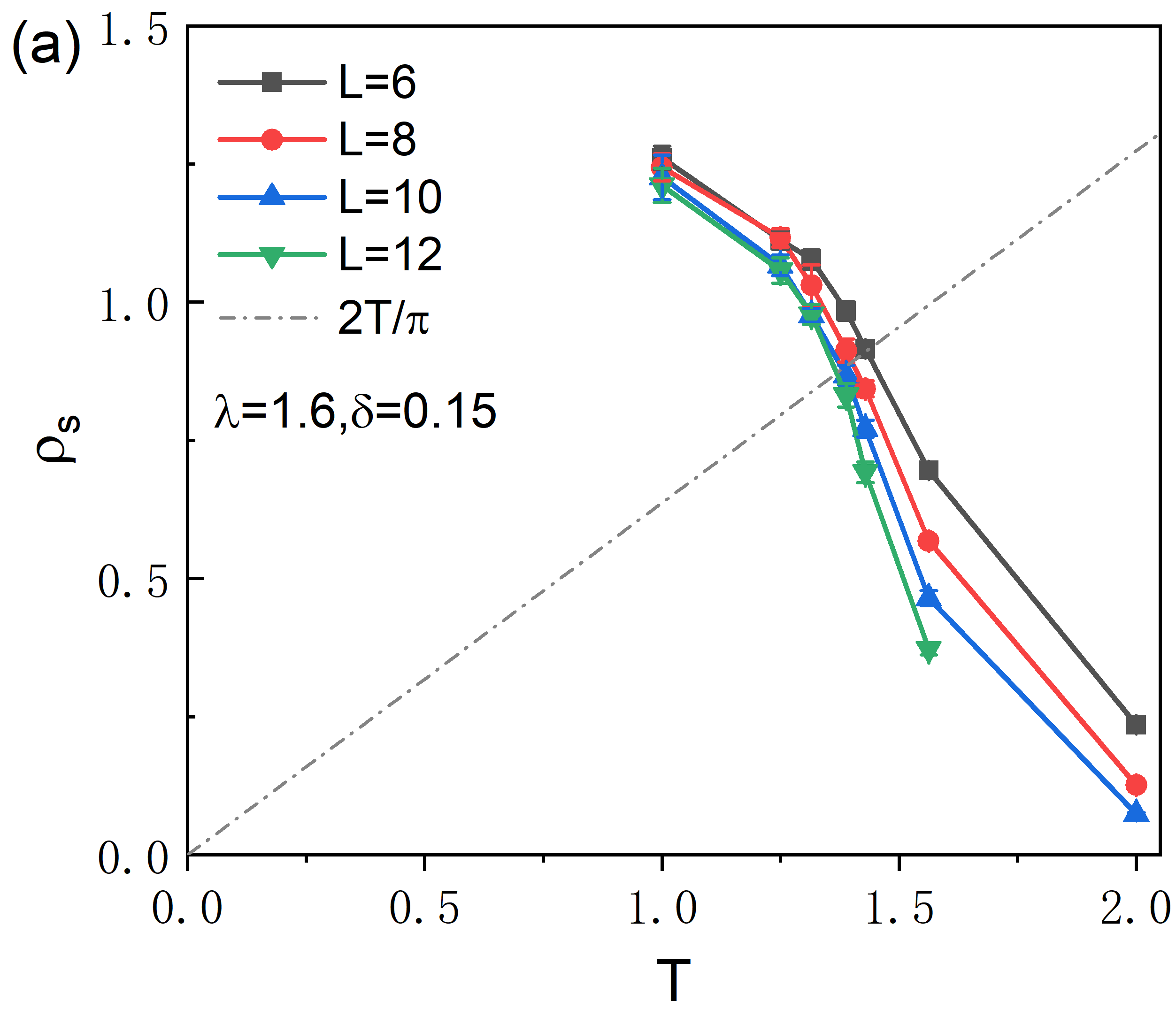}~~
\includegraphics[width= 0.46\linewidth]{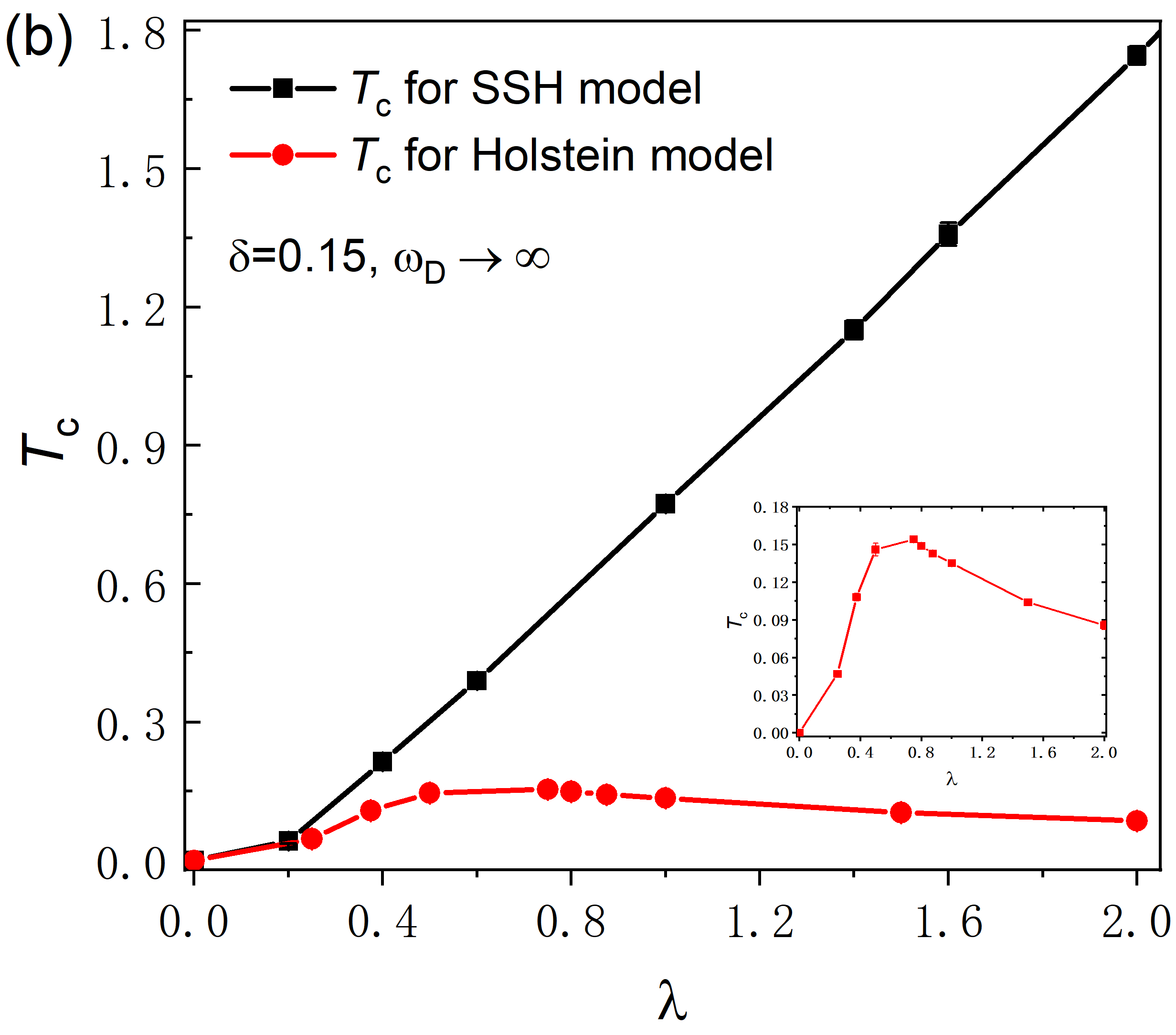}	
\caption{QMC results of superconductivity in the anti-adiabatic (AA) limit ($\omega_D=\infty$). (a) The superfluid stiffness $\rho_s$ as a function of $T$ for $\lambda=1.6$ and $\delta=0.15$. Lattice system size ranges from $L=6$ to $L=12$. The dashed line $2T/\pi$ is used to extract $T_c$ of the superconductivity (BKT) transition. (b) $T_c$ versus $\lambda$ for the SSH and Holstein model in the AA limit, respectively, with doping $\delta=0.15$. It is clear that $T_c$ of the SSH model increases linearly with $\lambda$ without bound in the strong coupling limit while $T_c$ of the Holstein model is suppressed with increasing $\lambda$ in the strong coupling regime (see the inset for $T_c$ of the Holstein model).}
\label{Fig2}
\end{figure}

Why can the SSH phonons mediate superconductivity with much higher $T_c$, as reliably shown from our QMC study? This is mainly because, distinct from Holstein EPC, large pair hopping amplitude can be effectively generated by the SSH phonons, which helps the phase coherence of the formed Cooper pairs and thus boosts superconducting $T_c$. For finite phonon frequency $\omega_D$, VBS instability appears only for sufficiently strong $\lambda$, which competes with SC and starts to suppress SC. And the maximum $T_c$ occurs around the quantum critical points of the model at half filling separating the AFM and VBS phases. Nonetheless, the maximum $T_{c,\textrm{SSH}}^{\textrm{max}}$ of the SSH model in this case is still considerably higher than that of the Holstein model for which bipolarons with large effective mass form for large $\lambda$ and $T_c$ is strongly suppressed mainly due to weak phase coherence of Cooper pairs. 
More strikingly, in the anti-adiabatic (AA) limit (i.e. $\omega_D\to \infty$), $T_c$ of the SSH model grows without bound as $\lambda$ increases, which is sharply distinct from the Holstein model in the same AA limit, as shown in \Fig{Fig2}. The higher-$T_c$ feature of the SSH model in parameter regime relevant to realistic materials indicates that it is promising to realize high-$T_c$ SC induced by SSH phonons which are dominant EPC in the materials.

\textbf{Model}: We consider the bond SSH model on the square lattice as described by the following Hamiltonian:
\bea\label{EqDefSSH}
&&H= -t\sum_{\avg{ij},\s}(c^\dagger_{i\s}c_{j\s}+\mathrm{h.c.}) -\mu\sum_{i\s}n_{i\s} \nonumber\\
&&~~~~~~ +\sum_{\avg{ij}}\frac{\hat{P}_{ij}^2}{2M}+\frac{K}{2}\hat{X}_{ij}^2 +g\sum_{\avg{ij},\s} \hat{X}_{ij}(c^\dagger_{i\s}c_{j\s}+\mathrm{h.c.}),~~~
\eea
where $c^\dagger_{i\s}$ creates an electron on site $i$ with spin polarization $\s=\uparrow,\downarrow$ and  $\hat{X}_{ij}$ ($\hat{P}_{ij}$) are displacement (momentum) operators of SSH phonons residing on the bond between nearest-neighbor (NN) sites $\avg{ij}$.
Here $t$ is the hopping amplitude of electrons between NN sites and $g$ denotes the coupling between electrons and phonons.
The SSH phonon frequency is $\omega_D=\sqrt{K/M}$.
The EPC strength is characterized by a dimensionless constant $\lambda\equiv \frac{4g^2/K}{W}$, where $W=8t$ is the characteristic band width of electron on square lattice. Hereafter, we set $t=1$ as energy unit and $K=1$ by appropriately redefining the phonon displacement fields $\hat{X}_{ij}$. By adjusting the chemical potential $\mu$, we can study the system away from half-filling. Due to the particle-hole symmetry in \Eq{EqDefSSH}, we will focus on the case of doping holes away from half filling.

The bond SSH model in \Eq{EqDefSSH} has attracted increasing interest recently \cite{Scalettar2021PRL,XC2021PRL, Cai2022PRBSSH,Assaad2022PRBSSH,Scalettar2022PRB}.
At half-filling, the model respects both spin and pseudospin SU(2) symmetry \cite{XC2021PRL}.
The pseudospin rotation transforms CDW order into on-site SC order \cite{XC2021PRL,Assaad2022PRBSSH}.
An additional particle-hole symmetry for spin-down electrons at half-filling, $c_{i\downarrow}\ra\inc{-1}^i c^\dagger_{i\downarrow}$, which transforms the spin SU(2) operators into the pseudospin SU(2) operators or vice versa, guarantees the degeneracy between the AFM ground state and the CDW/SC ground state.
The half-filled SSH model features spin or pseudospin AFM long-range order induced by SSH phonons within a large parameter regime of $\lambda$ and $\omega_D$, as shown in \Fig{Fig1}(a). 
For strong enough $\lambda$ or low enough $\omega_D$, valence bond solid (VBS) ordering becomes dominant, and the system undergoes a quantum phase transition between AFM and VBS insulating phases with increasing $\lambda$ or reducing $\omega_D$. 
The ground state phase diagram of \Eq{EqDefSSH} at half filling was obtained in Ref. \cite{XC2021PRL}, as depicted in \Fig{Fig1}(a).

In this Letter, we investigate possible SC emerging from doping the SSH model away from half filling, as it is widely believed that doping AFM could possibly lead to high-temperature SC. The SSH model in \Eq{EqDefSSH} is free from the sign problem at any filling and we can access the properties of doped SSH model with large system size and low temperature through numerically-exact QMC simulations \cite{Assaadnote,ZXLiQMCreview}.
Thus, we performed large-scale determinant QMC simulations to systematically study SC and other possible competing ordering in the doped SSH model in various parameter regimes.
As the system under study is 2D, we identify its superconducting $T_c$ using the relation $\rho_s(T\ra T_c^-)=2T_c/\pi$ for BKT transitions, where $\rho_s(T)$ is the superfluid stiffness that can be accurately extracted from QMC simulations. To gain further insight into the physics of SSH phonon mediated SC and decipher the underlying mechanism of its possible high transition temperature, we also study SC in the Holstein model at the same doping level for comparison; the phonon-related terms in the Holstein model reads $\sum_i (\hat P_i^2/2M_h+K_h \hat X_i^2/2)+g_h\sum_i \hat{X}_i(n_{i}-1)$ with $\omega_D=\sqrt{K_h/M_h}$ and $\lambda\equiv g_h^2/K_hW$. We compare the results of superconducting $T_c$ in doped SSH and Holstein models.  More details of the calculation are included in the Supplemental Material.

\textbf{Superconductivity at finite doping}:
We now explore SC in the lightly-doped SSH model. Hereafter the doping level is fixed to $\delta=0.15$ unless stated otherwise.
The physics of superconductivity emerging from lightly doped electron-phonon models can be probably best understood in the anti-adiabatic (AA) limit, where there is no retardation. So we first investigate SC in the AA limit and then study SC mediated by phonons with finite $\omega_D$ which is more relevant to real materials.

In the AA limit ($\omega_D\to \infty$), phonons can be integrated out exactly to generate instantaneous electron-electron interactions. The SSH model in the AA limit can be reduced to the following effective Hamiltonian:
\bea\label{EqDefAA}
H_{\mathrm{AA}} = -t\sum_{\avg{ij},\s}(c^\dagger_{i\s}c_{j\s}+\mathrm{h.c.})
+J\sum_{\avg{ij}}(\v{S}_i\cdot\v{S}_j +\tilde{\v{S}}_i\cdot\tilde{\v{S}}_j),
\eea
where $\v{S}_i=\frac{1}{2}c^\dagger_i\v{\s}c_i$ and $\tilde{\v{S}}_i=\frac{1}{2}\tilde{c}^\dagger_i\v{\s}\tilde{c}_i$ are spin and pseudospin operators on site $i$, respectively, with $c^\dagger_i=(c^\dagger_{i\uparrow}, c^\dagger_{i\downarrow})$, $\tilde{c}^\dagger_i=(c^\dagger_{i\uparrow}, \inc{-1}^i c_{i\downarrow})$ and $\vec{\sigma}=(\sigma_x,\sigma_y,\sigma_z)$, and $J=2g^2/K$ denotes the effective interaction strength mediated by SSH phonons.
To explicitly see the meaning of pseudospin interactions in $H_{AA}$, we rewrite the $J$ term on bond $\avg{ij}$ in \Eq{EqDefAA} as follows:
\bea
&&J(\v{S}_i\cdot\v{S}_j +\tilde{\v{S}}_i\cdot\tilde{\v{S}}_j)\nonumber\\
&&=J(\v{S}_i\cdot\v{S}_j+\frac{1}{4}n_in_j) -\frac{J}{2}\sum_{\avg{ij}}(c^\dagger_{i\uparrow}c^\dagger_{i\downarrow} c_{j\downarrow}c_{j\uparrow}+\mathrm{h.c.}),~~
\eea
which includes AF spin interaction, repulsive density interaction, and most importantly, the hopping of on-site pairs between NN sites with hopping amplitude $\frac{J}{2}$. This pair hopping, which is proportional to $J$, renders the possibility of high-$T_c$ SC upon doping away from half filling.

By performing numerically-exact QMC simulations of the SSH model in the AA limit $H_{\mathrm{AA}}$, we obtained the superfluid stiffness $\rho_s$ as a function of temperature $T$, as shown in \Fig{Fig2}(a), for $\lambda=0.4$.
The superconducting (BKT) transition temperature $T_c$ is extracted accurately from the intersection between the line $2T/\pi$ and the curves $\rho_s(T)$ for different system size $L$. 
We further obtained $T_c$ of the doped SSH model for various coupling strength $\lambda$, as shown in \Fig{Fig2}(b). As a comparison, we also calculated $T_c$ in the doped Holstein model in the AA limit, as shown by the red line in \Fig{Fig2}(b). $T_c$ of the doped Holstein model is suppressed at strong coupling regime of $\lambda$ due to the formation of heavy bipolarons, which suppresses the phase coherence of pairs. 
In contrast, $T_c$ of the lightly doped SSH model is remarkably higher, and exhibits an approximately linear relation with $\lambda$ in the strong coupling without bound.
In the SSH model, the superconducting $T_c$ can even exceed the hopping energy scale $t$ in the strong coupling.
The pair hopping process mediated by the SSH phonons is essential to delocalize the Copper pairs, rendering high temperature superconductivity possible. 

\begin{figure}[t]
\includegraphics[width= 0.49\linewidth]{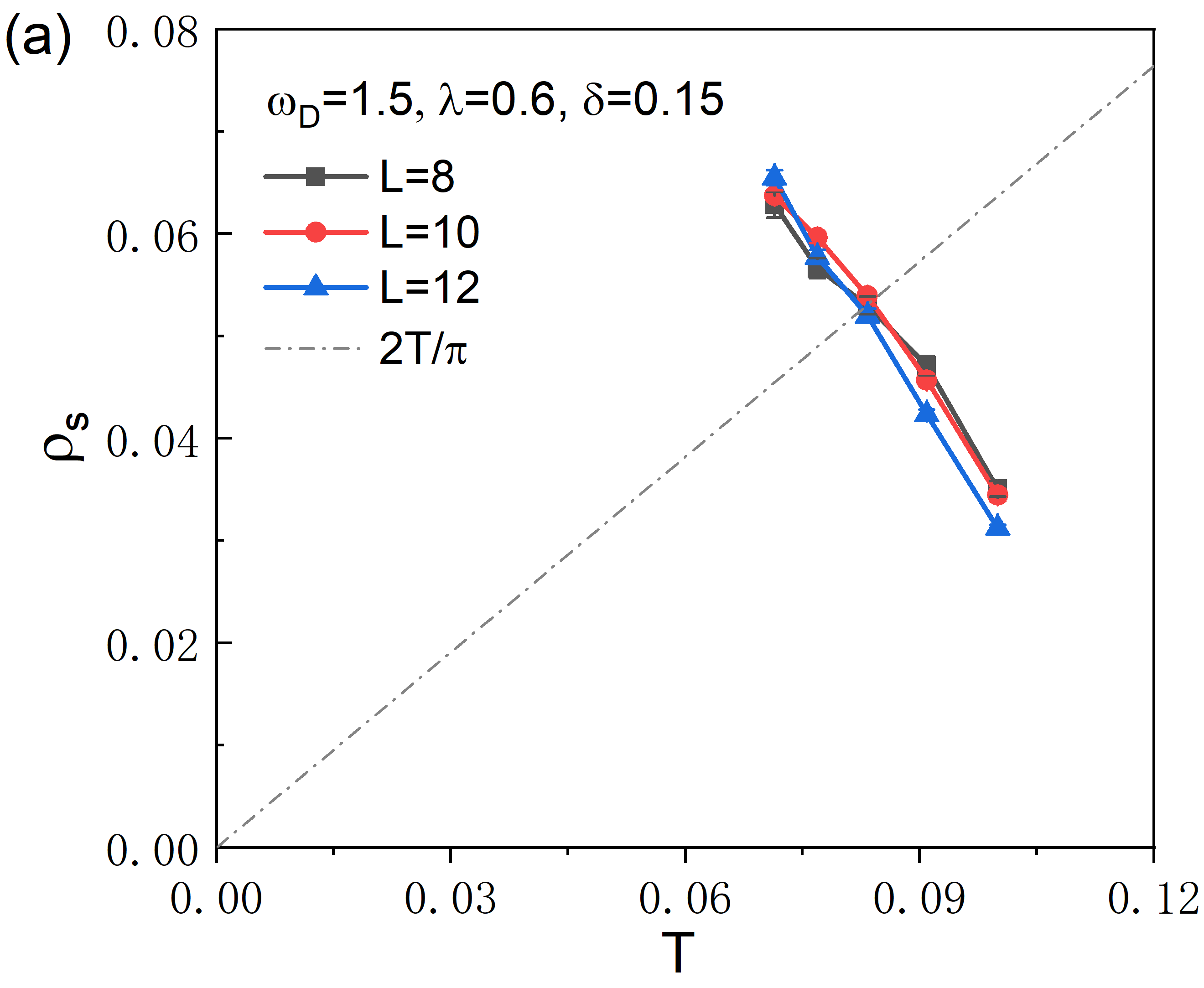}~~	
\includegraphics[width= 0.486\linewidth]{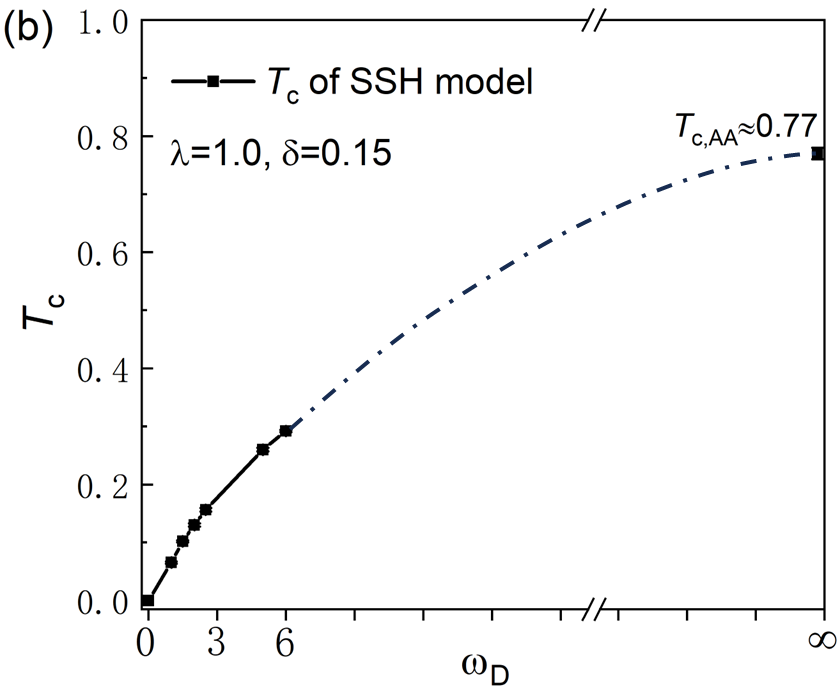}
\caption{(a) The QMC results of superfluid stiffness $\rho_s$ versus $T$ in the doped SSH model for $\omega_D=1.5$ and $\lambda=0.6$. Lattice size ranges from $L=8$ to $L=12$.  The intersection between dashed line $2T/\pi$ and curves of superfluid stiffness gives rise to the superconducting (BKT) transition temperature $T_c\approx 0.082$. (b)  $T_c$ of SSH model versus $\omega$ for fixed $\lambda=1.0$ and $\delta=0.15$. The value specified in the figures corresponds to the AA limit $\omega\ra\infty$. The black dashed line indicates the tendency towards AA limit as $\omega$ increases. }
\label{Fig3}
\end{figure}

We then investigate SC in the lightly-doped SSH model with a finite phonon frequency ($0<\omega_D<\infty$). 
At half filling, for a given finite $\omega_D$, AFM order is weakened and finally replaced by staggered VBS order with increasing $\lambda$, as illustrated in \Fig{Fig1}(a). After doping away from half filling, a possible high-$T_c$ SC may emerge for doping a AFM. Moreover, SC is also expected from doping a VBS, probably with lower $T_c$. To obtain its accurate $T_c$, we performed large-scale QMC simulations of the SSH models to calculate superfluid stiffness $\rho_s$, and extract $T_c$ from the universal behavior of $\rho_s$ at BKT transition point: $\rho_s(T\ra T_c^-)=2T_c/\pi$.
We first show the results of superfluid stiffness for doping a AFM ($\omega_D=1.5$ and $\lambda=0.6$) for different system sizes in \Fig{Fig3}(a), from which the SC transition temperature $T_c \approx 0.082$ is explicitly extracted. $T_c\approx 0.082t$ is about one percent of the band width $W=8t$, which is relatively high. For instance, $T_c\sim 300$ K if assuming $t=0.3$ eV. For a fixed $\lambda$, $T_c$ normally increases with $\omega_D$. Indeed, as shown in \Fig{Fig3}(b), $T_c$ increases linearly with $\omega_D$ for low frequency, and eventually saturates to $T_c$ of the AA limit.

\begin{figure}[t]
\subfigure{\label{fig4a}\includegraphics[width= 0.46\linewidth]{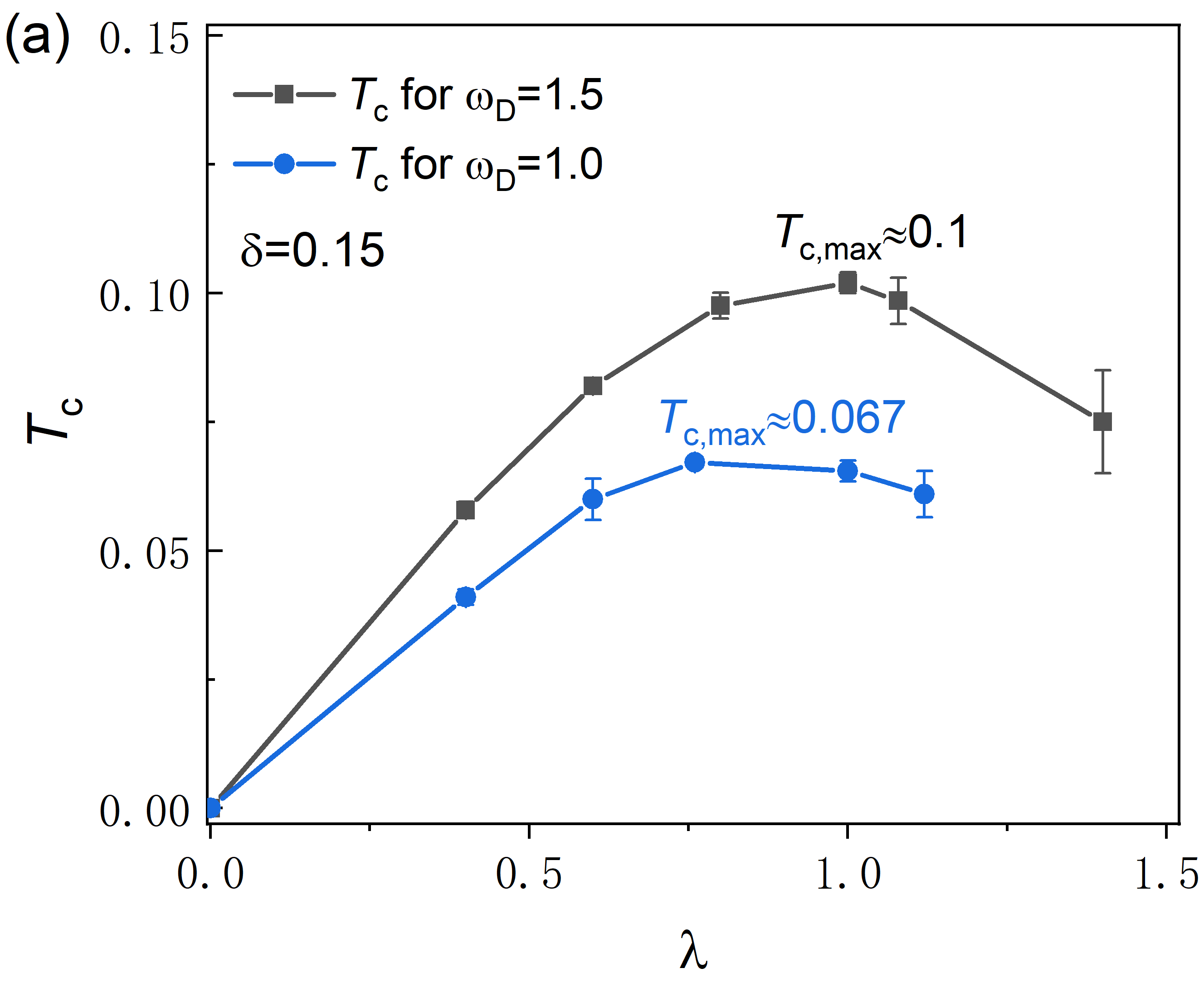}}~~
\subfigure{\label{fig4b}\includegraphics[width= 0.46\linewidth]{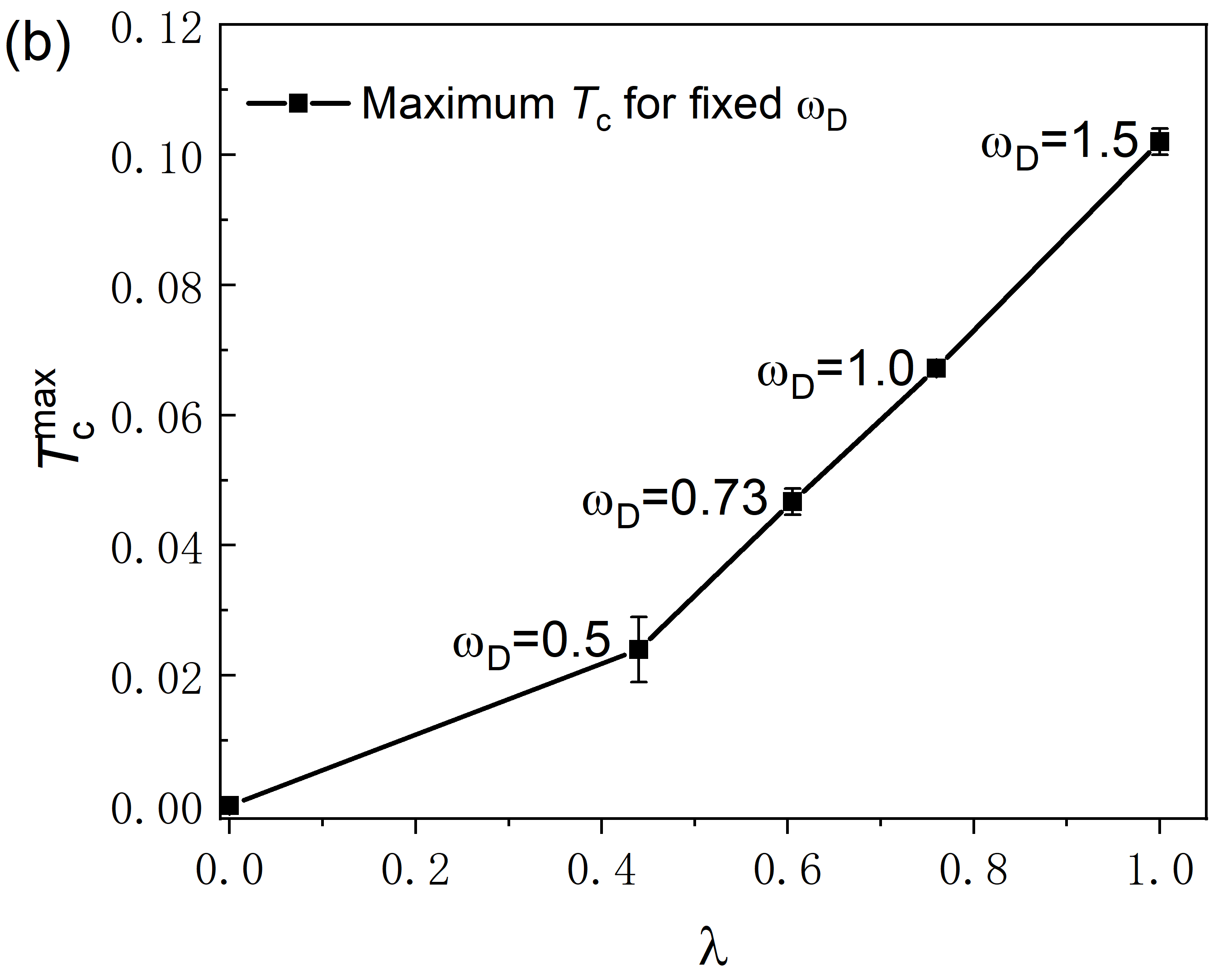}}
\caption{(a) $T_c$ in doped SSH model as the function of $\lambda$ for different $\omega$. The values specified in the figure correspond to the peaked $T_c$ for $\omega=1.5$ and $\omega=1$ respectively, the location of which coincide with the AFM-VBS phase boundary at half-filling as shown in \Fig{Fig1}(a). (b) The values of peaked $T_c$ versus $\lambda$ for $\delta=0.15$. The results are obtained along the phase boundary in \Fig{Fig1}(a) for each $\lambda$. The corresponding value of $\omega_D$ is marked for each data point.  }
\label{Fig4}
\end{figure}

We further investigated the behaviour of $T_c$ with varying $\lambda$ by QMC. In \Fig{Fig1}(b), we present the dependence of $T_c$ on $\lambda$ for the SSH model with fixed phonon frequency $\omega_D=1.5$. For $\omega_D=1.5$, $T_c(\lambda)$ exhibits a pronounced dome-like behaviour, displaying a peak at the optimal value of $\lambda = \lambda_o \approx 1.0$. More intriguingly, a comparison with the quantum phase diagram at half filling in \Fig{Fig1}(a) suggests that the optimal value of $\lambda=\lambda_o$ for the peak $T_c$ coincides with the quantum critical point $\lambda=\lambda_c \approx 1.0$ between the AFM and VBS phases at half filling. Namely $\lambda_o\approx \lambda_c\approx 1.0$ for $\omega_D=1.5$.
Interestingly, $\lambda_o \approx \lambda_c\approx 0.75$ is further observed for another phonon frequency $\omega_D=1.0$, as shown in \Fig{fig4a}. We plot the optimal $T_c$ reached at $\lambda=\lambda_o$ in the SSH model for various $\omega_D$ in \Fig{fig4b}, each of which is approximately located on AFM-VBS phase boundary $\lambda_c$ at half filling as shown in \Fig{Fig1}(a). It suggests that highest $T_c$ can be achieved in the doped SSH model by turning $\lambda$ around quantum critical points of the parent model. 

Our observation that $\lambda_o \approx \lambda_c$ clearly implies that SC behaves qualitatively different between lightly-doping a AFM and lightly-doping a VBS. 
$T_c$ of SC emerging from a lightly-doped AFM phase increases with $\lambda$; and conversely, it is suppressed with increasing $\lambda$ for a lightly-doped VBS phase.
In the parent AFM phase at half filling, the pseudospin AFM (namely CDW/SC) state is actually degenerate with the AFM state owing to the special particle-hole symmetry at half filling of the SSH model. 
Upon doping, the degeneracy is broken and the ground state becomes a superconductor with a finite $T_c$ even when the doping is infinitesimally small.  
With increasing $\lambda$, the effective pair hopping amplitude is enhanced, increasing its superconducting $T_c$ as long as the parent ground state is in the AFM phase.

In contrast, the VBS order appearing at large $\lambda$ is robust such that infinitesimal doping cannot turn the system into a superconductor; namely there exists a critical doping level at which a quantum phase transition occurs between the VBS phase and a superconducting phase.
For light doping from the parent VBS phase of the SSH model, the bond bipolaron is formed for $\lambda$ around the half-filling quantum critical point $\lambda_c$ and the effective mass of bipolarons tends to increase with increasing $\lambda$.
Hence, for the SC from lightly doped VBS phase, $T_c$ is reduced with increasing $\lambda$ due to the suppression of phase coherence of pairs by the formation of heavy bond bipolarons. 
For $\lambda$ sufficiently larger than $\lambda_c$, strong bond ordering with vanishing superconductivity is observed in our QMC simulation, as expected. 
The details are included in Supplemental Materials.
The competition between bond ordering and SC is a crucial factor accounting for the suppression of $T_c$ at very strong EPC regime. If the half-filling ground state were a resonating valence bond (RVB), namely a quantum spin liquid, such suppression of $T_c$ may not occur and a high $T_c$ could emerge, as proposed in early days of cuprate superconductivity \cite{Anderson1987RVB, Anderson2004-vanillaRVB, Rokhsar-Kivelson1988PRL, Wen2006RMP}.

To further understand the physics of high $T_c$ obtained in doped SSH model, we performed QMC simulations of the doped Holstein model for fixing phonon frequency $\omega_D=1.5$ and varying its EPC strength $\lambda$ and compared it with the results of the SSH model with the same $\omega_D$. 
For the Holstein model with $\omega_D=1.5$ and doping $\delta=0.15$, SC is not observed even at the lowest temperature reached in our simulations, implying that $T_c$ of the Holstein model is quite low.
The lowest temperature in the QMC simulations we reached for the Holstein model can be considered as its upper bound of $T_c$.
In \Fig{Fig1}(b) we present the upper bound of $T_c$ of the doped Holstein model for $\omega_D=1.5$.
The superconducting $T_c$ in the doped Holstein model is extremely low $T_c<0.01$ for $\omega_D=1.5$. 
In contrast to Holstein EPC, the doped SSH model features a high-$T_c$ SC. 
For the SSH model, the maximum $T_c\approx 0.1$ is obtained at $\lambda=\lambda_o\approx 1.0$.  
Then, we can estimate that the ratio between the maximum $T_c$ of the two models $T^{\mathrm{max}}_{c,\mathrm{SSH}}/T^{\mathrm{max}}_{c,\mathrm{Holstein}}\geq 10$ for $\omega_D=1.5$. 
The remarkable high-$T_c$ in the SSH model is attributed to the pair hopping induced by SSH phonons, which is effectively absent in the Holstein model. The pair hopping processes (although retarded for finite $\omega_D$) arising from the SSH phonons can effectively suppress the formation of bond bipolaron with large effective mass, hence boosting phase coherence temperature of the resulting superconductivity. 

\textbf{Conclusions and discussions}: 
In this work, from numerically-exact QMC simulations, we unambiguously demonstrated that a remarkably high $T_c$ can be achieved in the doped SSH model for large parameter regime, in stark contrast with the doped Holstein model with relatively low $T_c$. 
In particular, in the AA limit ($\omega_D=\infty$) where the phonon-mediated interaction is instantaneous without retardation, the effective pair hopping interaction scales linearly with $\lambda$, thus enhancing $T_c$ that without bound.
For finite $\omega_D$, the doped SSH model features a dome-like $T_c$ with increasing $\lambda$, with the optimal $T_c$ located around the quantum critical point between the AFM and VBS phases at half filling.
Such results suggest that in the VBS phase with relatively large $\lambda$, pair hopping is suppressed due to the formation of bipolarons with large effective mass arising from strong EPC. Our state-of-the-art numerical simulations on the lightly doped SSH model implies that strong pair hopping mediated by SSH phonons is crucial in realizing large phase coherence of Cooper pairing and high $T_c$ of superconductivity. Such strong pair hopping is more likely to be realized in doping a AFM or a RVB.

We emphasize that our simulation is qualitatively different from the studies of bipolaronic superconductor in dilute limit \cite{Millis2023PRX}. In our study, we investigate the system at a macroscopic filling, in which various orderings such as VBS are present and intertwined with SC. The model with macroscopic filling is more relevant to realistic materials, enabling us to identify accurate superconducting (BKT) transition temperature through the scaling behaviour of superfluid stiffness.

Although we have focused on the simplest SSH model, we believe that the conclusion of achieving higher $T_c$ in such model with large pair hopping can qualitatively apply to more generalized bond-type EPC, for instance the $B_{1g}$ type EPC studied in cuprates \cite{DevereauxCuprate2010PRB,Lanzara2001Cuprate,He2018Science}.
In all the parameter regime of $\lambda$ and $\omega_D$ we have simulated, our simulations show that the $T_c$ in the SSH model is significantly higher compared with that of the Holstein-type EPC, thus pointing out a promising direction to searching for high-$T_c$ superconductors in quantum materials with dominant SSH type EPC.

\textit{Acknowledgement}: We would like to thank Fu-Chun Zhang for helpful discussions. This work is supported in part by the NSFC under Grant No. 11825404 and the MOSTC Grant No. 2021YFA1400100 (HY). Z.X.L acknowledges support from the start-up grant of IOP-CAS. 

\bibliography{doped_SSH}

\begin{appendices}
\widetext
\begin{center}
    \section{Supplemental Material}
\end{center}
\setcounter{equation}{0}
\setcounter{figure}{0}
\setcounter{table}{0}
\makeatletter
\renewcommand{\theequation}{S\arabic{equation}}
\renewcommand{\thefigure}{S\arabic{figure}}
\renewcommand{\thesubfigure}{\thefigure(\alph{subfigure})}
\renewcommand{\bibnumfmt}[1]{[S#1]}

\subsection{A. Details of determinant quantum Monte Carlo}
In the present paper, we evaluate the superconducting transition temperature in both doped SSH and Holstein model by employing numerically exact determinant quantum Monte Carlo (DQMC). For generic electron-phonon coupled systems, the partition function is formulated as
\bea\label{EqDefPartition}
Z=\int \Dif X\ \E{-S_B[X]}\,\Tr_\psi\Inc{\E{-S_F[\bar{\psi},\psi,X]}}
\eea
where $X$ denotes the phonon field and $\psi$ the electron field. The algorithm treats the fermionic trace within the integrand by Suzuki-Trotter decomposition
\bea\label{EqFermionTrace}
\Tr_\psi\Inc{\E{-S_F}}=\Tr_\psi\Inc{\prod^{L}_{l=1}\hat{U}_l}+\mathcal{O}(\Delta\tau^2)
\eea
with discrete time slices defined as $\beta=L\Delta\tau$, $\tau=l\Delta\tau$, and the imaginary time propagator
\bea\label{EqDefPropagator}
\hat{U}_l=\E{-\Delta\tau\psi^\dagger K\psi}\E{-\Delta\tau\psi^\dagger V_l \psi}
\eea
where $K$ is the electron kinetic matrix and $V_l$ is the electron-phonon coupled matrix that depends explicitly on the phonon fields on time slice $\tau=l\Delta\tau$. For Holstein model, $\inc{V_l}_{ij,\s\s^\prime}=\delta_{ij}\delta_{\s\s^\prime}g X_{i,l}$, and for SSH model, $\inc{V_l}_{ij,\s\s^\prime}=\delta_{\s\s^\prime}\delta_{\avg{ij}}gX_{\avg{ij}}$. Integrating out the fermions explicitly after the imaginary time discretization yields
\bea\label{EqDefDet}
\Tr_\psi\Inc{\prod^{L}_{l=1}\hat{U}_l}=\Det{\Inc{1+\prod^L_{l=1}B_l}}=\Det\ G^{-1}_X
\eea
with $B_l=\E{-\Delta\tau K}\E{-\Delta\tau V_l}$. Note that for both SSH and Holstein model the action is degenerate with spin indices, and the propagator $B_l$ is real in each spin sector. Thus in practice we calculate the Green function $G_X$ in a single spin sector, denoted as $\tilde{G}_X$, and each phonon field configuration $X=\{X_{i,l}\}$ is sampled by importance according to the weight function
\bea\label{EqWeight}
W[X]=\E{-S_B[X]}\inc{\Det\ \tilde{G}^{-1}_X}^2
\eea
in which the fermionic sign problem is automatically avoided.

In the AA limit the SSH model is reduced to
\bea\label{EqDefAA2}
H_{\mathrm{AA}}=-t\sum_{\avg{ij}}\inc{c^\dagger_{i\s}c_{j\s}+\mathrm{h.c.}}-\frac{J}{4}\sum_{\avg{ij}}\inc{\sum_{\s}c^\dagger_{i\s}c_{j\s}+\mathrm{h.c.}}^2
\eea
which is equivalent to \Eq{EqDefAA} in the main text. In order to enable DQMC algorithm on \Eq{EqDefAA2}, an additional Hubbard-Stratonovich (H-S) transformation on the instantaneous four-fermion interaction is necessary. We implement a four-fold discrete H-S transformation following the identity\cite{Assaadnote}
\bea\label{EqDefHS}
\E{\Delta\tau\lambda \hat{A}^2}=\sum_{s=\pm 1,\pm 2}\gamma_s\E{\sqrt{\Delta\tau\lambda}\eta_s\hat{A}}+
\mathcal{O}(\Delta\tau^4)
\eea
with $\lambda$ the coupling coefficient and $\hat{A}$ an arbitrary one-body operator. The auxiliary fields $\gamma,\eta$ are defined as
\bea\label{EqAuxField}
\begin{aligned}
    \gamma_{\pm 1} &= 1+\sqrt{6}/3 \\
    \gamma_{\pm 2} &= 1-\sqrt{6}/3 \\
    \eta_{\pm 1} &= \pm \sqrt{2\inc{3-\sqrt{6}}} \\
    \eta_{\pm 2} &= \pm \sqrt{2\inc{3+\sqrt{6}}}
\end{aligned}
\eea
Henceforth we retrieve the same formalism of Trotter decomposition as in \Eq{EqFermionTrace}. \Eq{EqDefAA2} is also free of sign problem in this formalism.

In our practical QMC simulation, we set the Trotter time slice $\Delta\tau=0.1/t$. Convergence of the discretization has been checked for different system sizes $L$ by comparing with smaller $\Delta\tau$. Lattice system size ranges from $L=6$ to $L=12$ for SSH and Holstein model. For specific case in SSH model at AA limit, the largest size may push to $L=14$, as specified in section D.

\subsection{B. Calculation of superfluid stiffness}
\begin{figure}[t]
\includegraphics[width= 0.4\linewidth]{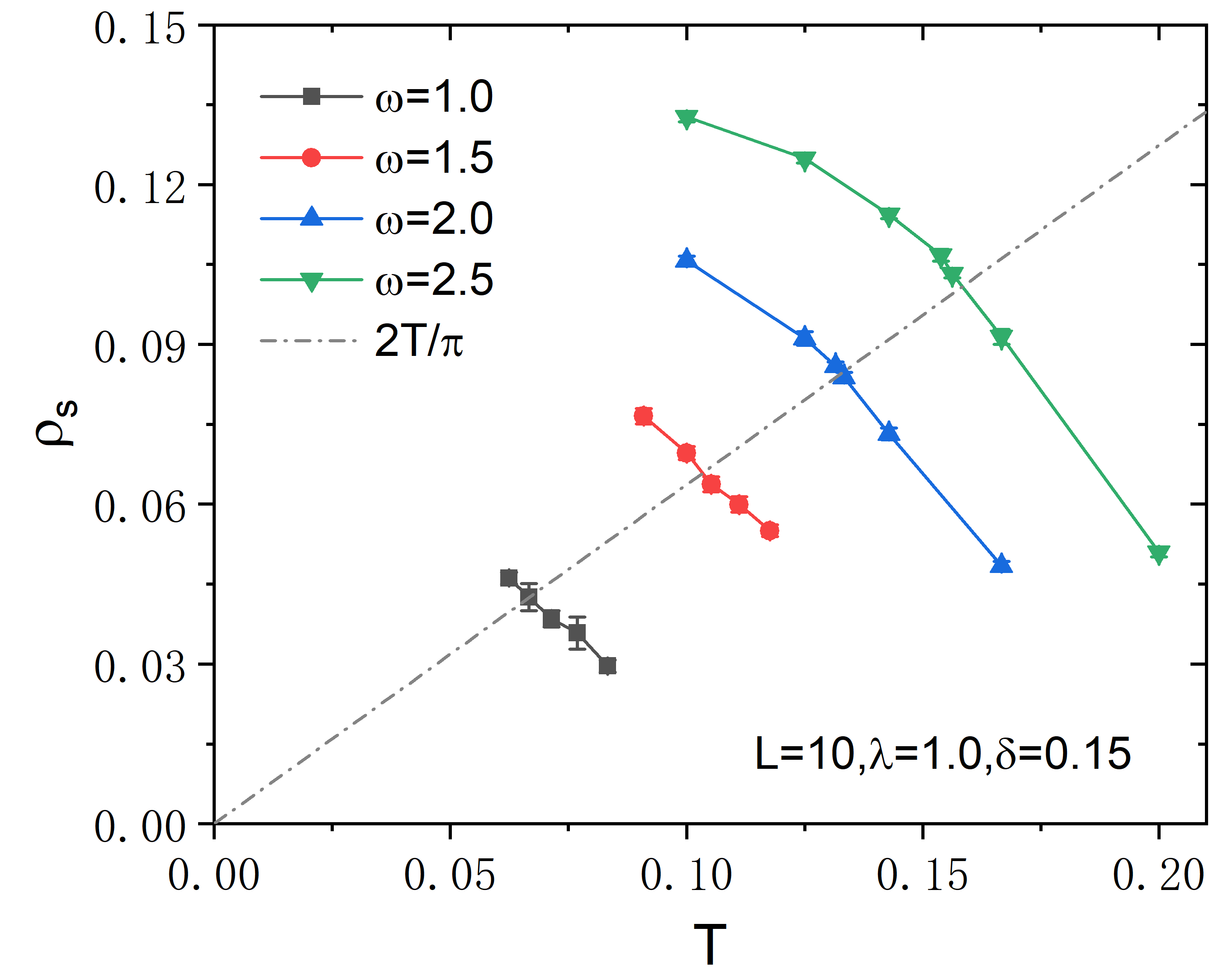}
\caption{QMC results of $\rho_s$ as the function of $T$ in SSH model for fixed $L=10$, $\lambda=1.0$ and $\delta=0.15$  . The grey dashed line is the guideline $2T/\pi$ for identifying $T_c$ by the intersection with $\rho_s(T)$ for different phonon frequency $\omega_D$. }
\label{FigS1}
\end{figure}

In our work we extract superconducting $T_c$ from the universal behavior of superfluid stiffness $\rho_s(T)$ for BKT transition: $\rho_s(T\ra T_c^-)=2T_c/\pi$. To evaluate $\rho_s$ in fermionic lattice systems, we follow the derivation specified in\cite{SCZhang_PRL_SFdensity,SCZhang_PRB_SFdensity} and relate $\rho_s$ with current-current correlators in the form as
\bea\label{EqDefSF}
\rho_s=\frac{1}{4}\Inc{-K_x-\Lambda_{xx}(q_x=0,q_y\ra 0,\omega_n=0)}
\eea
where $K_x$ is the diamagnetic current response along lattice $x$-direction, with the assumption of the external gauge field along $x$-direction. $\Lambda_{xx}(\v{q},\omega_n)$ is the paramagnetic current correlator defined as
\bea\label{EqDefParaCurr}
\Lambda_{xx}(\v{q},\omega_n)=\frac{1}{L^d}\sum_{i,j}\int^\beta_0\,\dif\tau\,\E{\imth\omega_n\tau}\E{-\imth\v{q}\cdot\inc{\v{R}_i-\v{R}_j}}\,\avg{j_x(\v{R}_i,\tau)j_x(\v{R}_j,0)}
\eea
The U(1) gauge invariance guarantees that the longitudal paramagnetic response in the long-wavelength limit is strictly related to the diamagnetic response
\bea\label{EqGauge}
\Lambda_{xx}(q_x\ra 0,q_y=0,\omega=0)+K_x=0
\eea
Thus the superfluid stiffness can also be written in terms of the longitudal and transverse response as
\bea\label{EqDefSF2}
\rho_s=\frac{1}{4}\inc{\Lambda^L-\Lambda^T}
\eea

In our QMC simulation we calculate the imaginary time current correlators to estimate $\rho_s$ in use of \Eq{EqDefSF2} and \Eq{EqDefParaCurr}. The paramagnetic current operator in Holstein model follows the usual definition
\bea\label{EqHolsteinCurrent}
j_x^{\mathrm{Holstein}}(\v{R}_i)=\imth t\sum_\s \inc{c^\dagger_{i+x,\s}c_{i,\s}-c^\dagger_{i,\s}c_{i+x,\s}}
\eea
while in SSH model the paramagnetic current is modulated by lattice distortions:
\bea\label{EqSSHCurrent}
j_x^{\mathrm{SSH}}(\v{R}_i)=\imth \inc{t-g\hat{X}_i}\sum_\s \inc{c^\dagger_{i+x,\s}c_{i,\s}-c^\dagger_{i,\s}c_{i+x,\s}}
\eea
For SSH model in AA limit, the singlet pairs $\Delta^\dagger_i=c^\dagger_{i\uparrow}c^\dagger_{i\downarrow}$ are also carriers for charge current in the presence of pair hopping interaction:
\bea\label{EqAACurrent}
j_x^{\mathrm{AA}}(\v{R}_i)=\imth t\sum_{\s}\inc{c^\dagger_{i+x,\s}c_{i,\s}-c^\dagger_{i,\s}c_{i+x,\s}}+\imth J\inc{\Delta^\dagger_{i+x}\Delta_i-\Delta^\dagger_i\Delta_{i+x}}
\eea
The finite size effect on estimating $\rho_s$ is controllable. In practice $\rho_s(T)$ extracted from QMC simulation is concentrated for different system size in the vicinity around $T_c$, enabling us to identify $T_c$ accurately with system size up to $L=12$. Examples have been provided in \Fig{Fig2}(a) for SSH AA limit and \Fig{Fig3}(a) for SSH model with $\omega_D=1.5$ in the main text. Here we present the results of $\rho_s(T)$ in SSH model for various phonon frequency with fixed $\lambda=1.0$ and $L=10$, as shown in \Fig{FigS1}, indicating the dramatic growth of $T_c$ with $\omega_D$.

\begin{figure}[t]
\subfigure{\label{figS2a}\includegraphics[width= 0.412\linewidth]{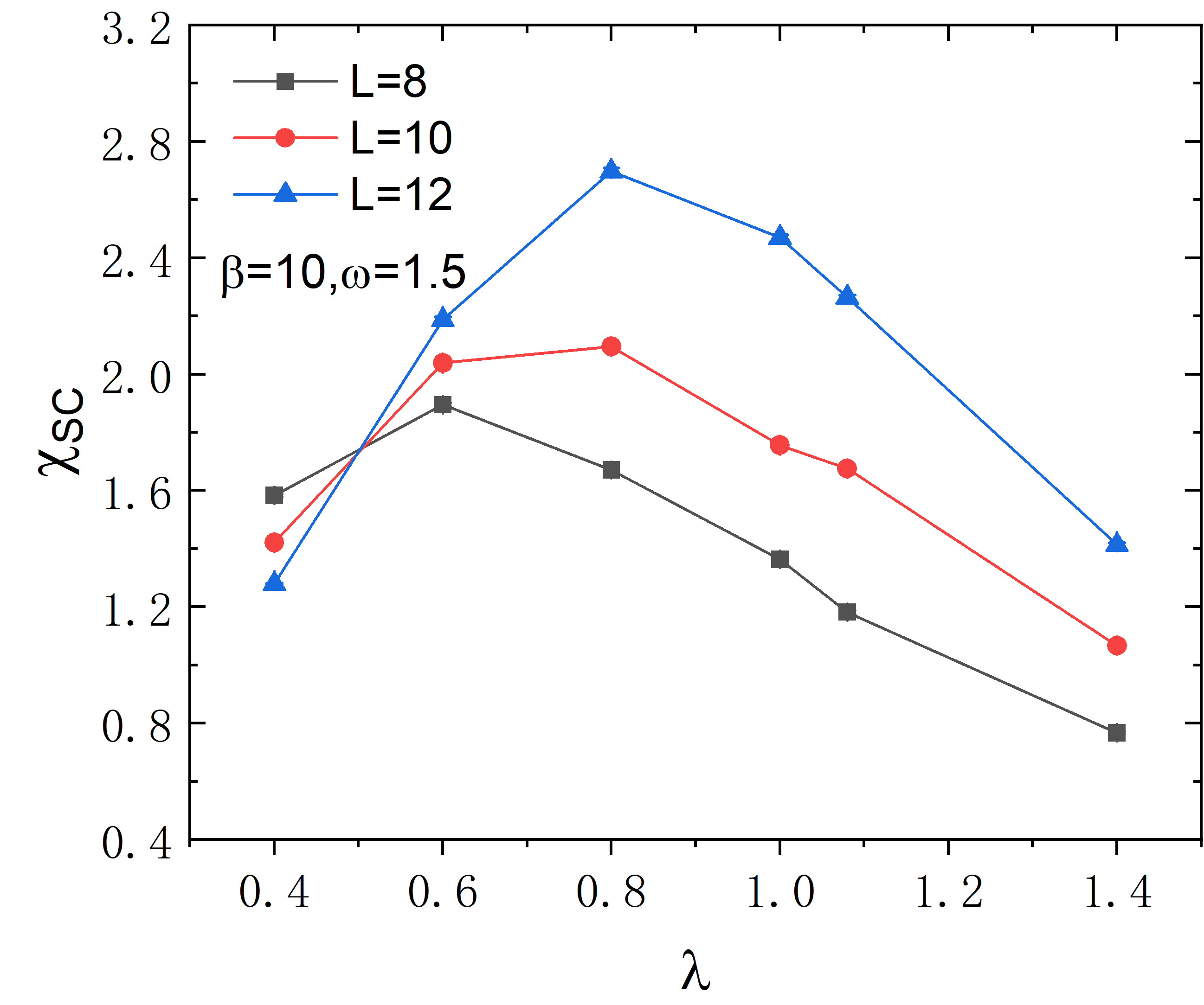}}~~~~	
\subfigure{\label{figS2b}\includegraphics[width= 0.408\linewidth]{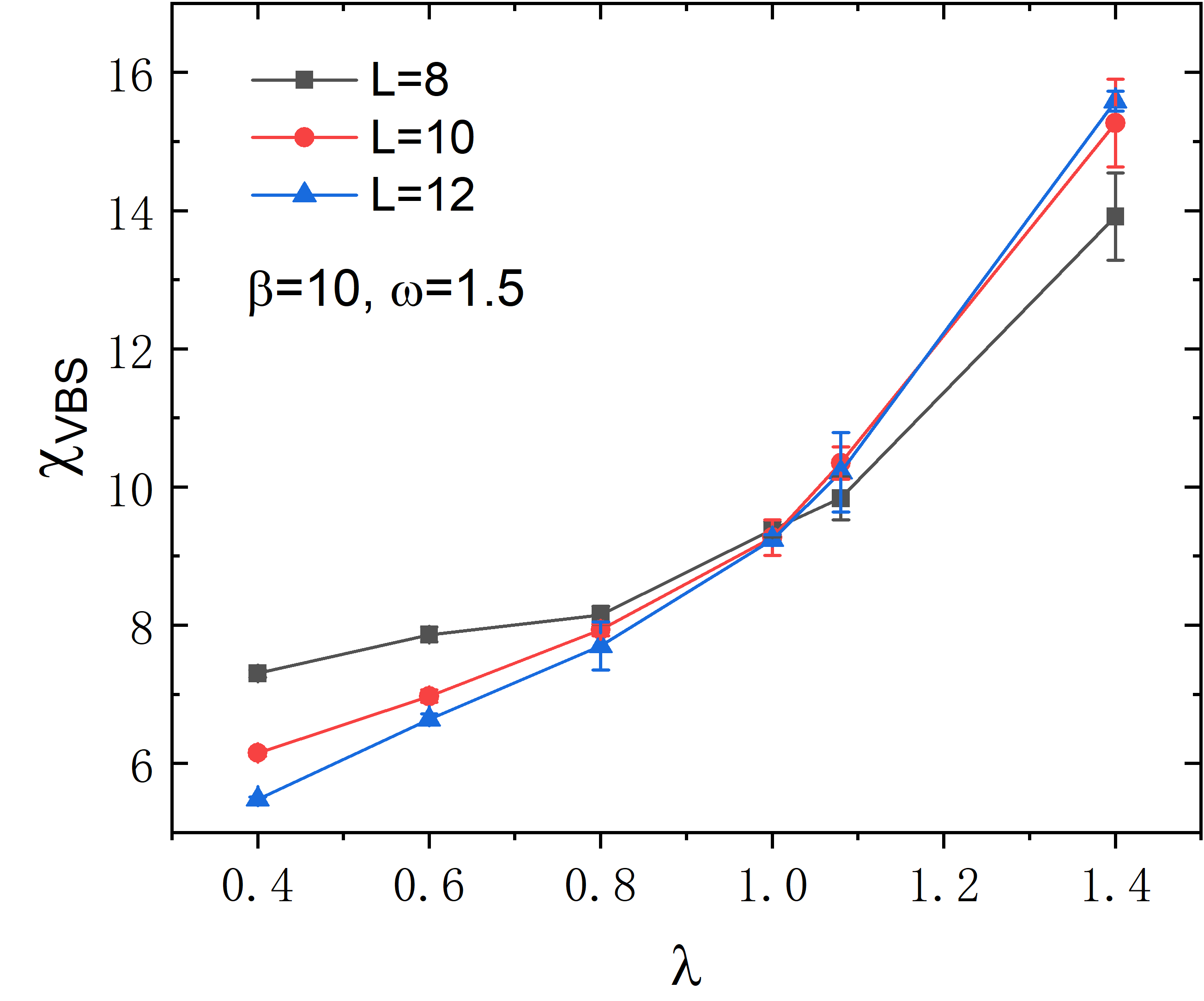}}
\caption{(a) The on-site pairing susceptibility $\chi_\mathrm{SC}$ and (b) the VBS susceptibility $\chi_\mathrm{VBS}$ in doped SSH model as the function of $\lambda$ for fixed $\beta=10$, $\omega_D=1.5$, and $\delta=0.15$. The largest system size is $L=12$. }
\label{FigS2}
\end{figure}

\begin{figure}[t]
\subfigure{\label{figS3a}\includegraphics[width= 0.415\linewidth]{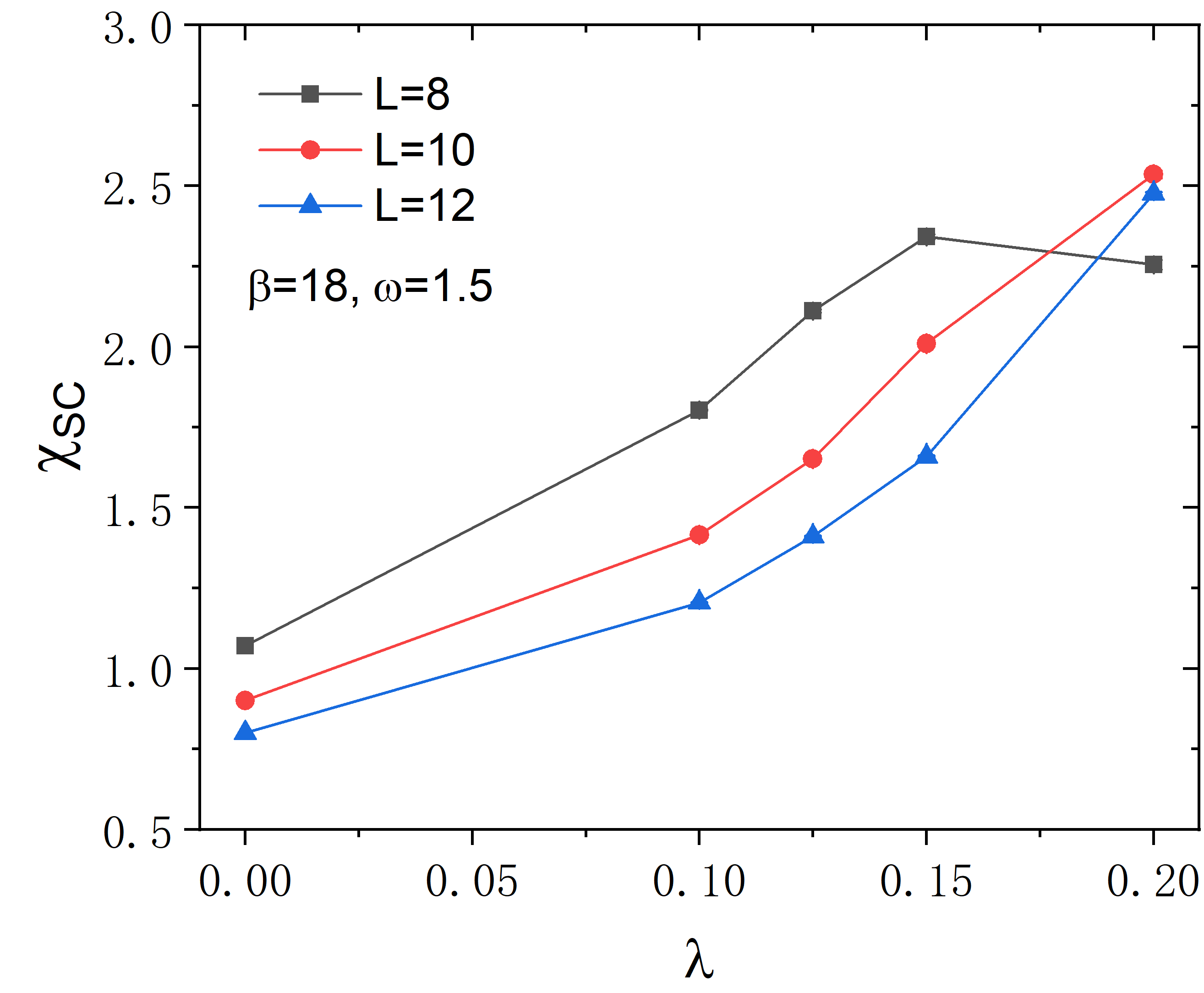}}~~~~	
\subfigure{\label{figS3b}\includegraphics[width= 0.4\linewidth]{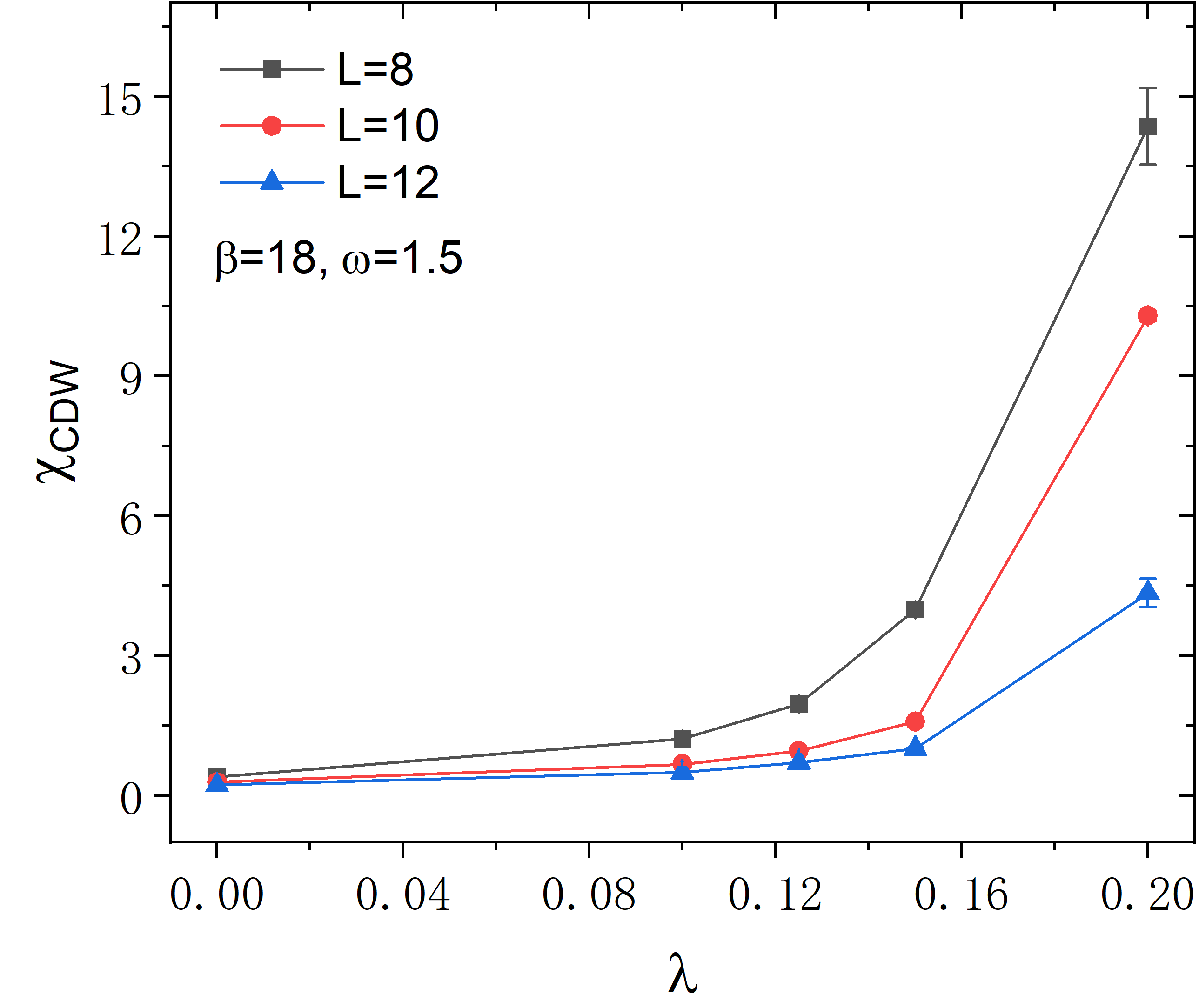}}
\caption{(a) The on-site pairing susceptibility $\chi_\mathrm{SC}$ and (b) the CDW susceptibility $\chi_\mathrm{CDW}$ in doped Holstein model as the function of $\lambda$ for fixed $\beta=18$, $\omega_D=1.5$, and $\delta=0.15$. The largest system size is $L=12$.}
\label{FigS3}
\end{figure}

\subsection{C. Lattice instability in SSH and Holstein model}
At strong coupling regime, superconductivity in both SSH and Holstein model suffers from the competition with other lattice instabilities. In SSH model, the bond-Peierls instability is the major obstacle to achieving high-$T_c$ SC. In \Fig{FigS2} we show both pairing and VBS susceptibilities as the function of $\lambda$ with the fixed temperature $\beta=10$ narrowly above superconducting $T_c$. The pairing susceptibility is defined as
\bea\label{EqDefSuscPair}
\chi_\mathrm{SC}=\frac{1}{L^d}\sum_{i,j}\int^\beta_0\,\dif\tau\,\avg{\hat{\Delta}^\dagger_i(\tau)\hat{\Delta}_j(0)}
\eea
with on-site pairing field $\hat{\Delta}_i=c_{i\downarrow}c_{i\uparrow}$. The VBS susceptibility is defined as
\bea\label{EqDefSuscVBS}
\chi_\mathrm{VBS}(\v{Q})=\frac{1}{L^d}\sum_{i,j}\int^\beta_0\,\dif\tau\,\E{-\imth\v{Q}\cdot\inc{\v{R}_i-\v{R}_j}}\avg{\inc{\hat{B}_{i,x}(\tau)+\imth\hat{B}_{i,y}(\tau)}\inc{\hat{B}_{j,x}(0)-\imth\hat{B}_{j,y}(0)}}
\eea
with bond kinetic operator $\hat{B}_{i,\delta}=\sum_\s c^\dagger_{i,\s}c_{i+\delta,\s}+\mathrm{h.c.}$ along lattice direction $\delta=x,y$. Ordering wave-vector for staggered-VBS is $\v{Q}=(\pi,\pi)$. As illustrated in \Fig{FigS2}(a), pairing susceptibility is boosted with $\lambda$ at weak coupling regime and decays at strong coupling. For fixed phonon frequency $\omega_D=1.5$, the peak location of $\chi_{\mathrm{SC}}$ moves towards the AFM-VBS phase boundary $\lambda_c\approx1.0$ at half-filling as the lattice system size $L$ enlarges. Meanwhile, the VBS susceptibility depicted in \Fig{FigS2}(b) enhances dramatically with $\lambda$ for all system size, especially in the regime $\lambda>1.0$ where the parent insulating phase possesses VBS long-range order at half-filling. The enhancement of $\chi_{\mathrm{VBS}}$ suggests that bond order instability driven by strong EPC is a dominant factor that suppresses superconducting $T_c$ even if the long-range VBS order is absent at light doping level.

Similar features occur in doped Holstein model. In \Fig{FigS3} we compare the tendency of pairing and CDW susceptibility as the function of $\lambda$, where CDW susceptibility is defined as
\bea\label{EqDefSuscCDW}
\chi_\mathrm{CDW}(\v{Q})=\frac{1}{L^d}\sum_{i,j}\int^\beta_0\,\dif\tau\,\E{-\imth\v{Q}\cdot\inc{\v{R}_i-\v{R}_j}}\avg{\inc{n_i(\tau)-\frac{1}{2}}\inc{n_j(\tau)-\frac{1}{2}}}
\eea
where the ordering wave-vector for CDW on square lattice is $\v{Q}=(\pi,\pi)$. In lightly doped Holstein model, $\chi_\mathrm{SC}$ and $\chi_\mathrm{CDW}$ enhances simultaneously with $\lambda$. Particularly, the increase of $\chi_\mathrm{CDW}$ is more significant for $\lambda>0.16$, suggesting the strong competition between SC and CDW instability in strong coupling regime. Nevertheless, the superconducting $T_c$ is even lower than the lowest temperature $T_c<0.02$ detectable in our QMC simulation, as mentioned in the main text. The effective pair hopping process, which is absent in Holstein model, is majorly responsible for the notable high $T_c$ in SSH model.

\begin{figure}[t]
\subfigure{\label{figS4a}\includegraphics[width= 0.40\linewidth]{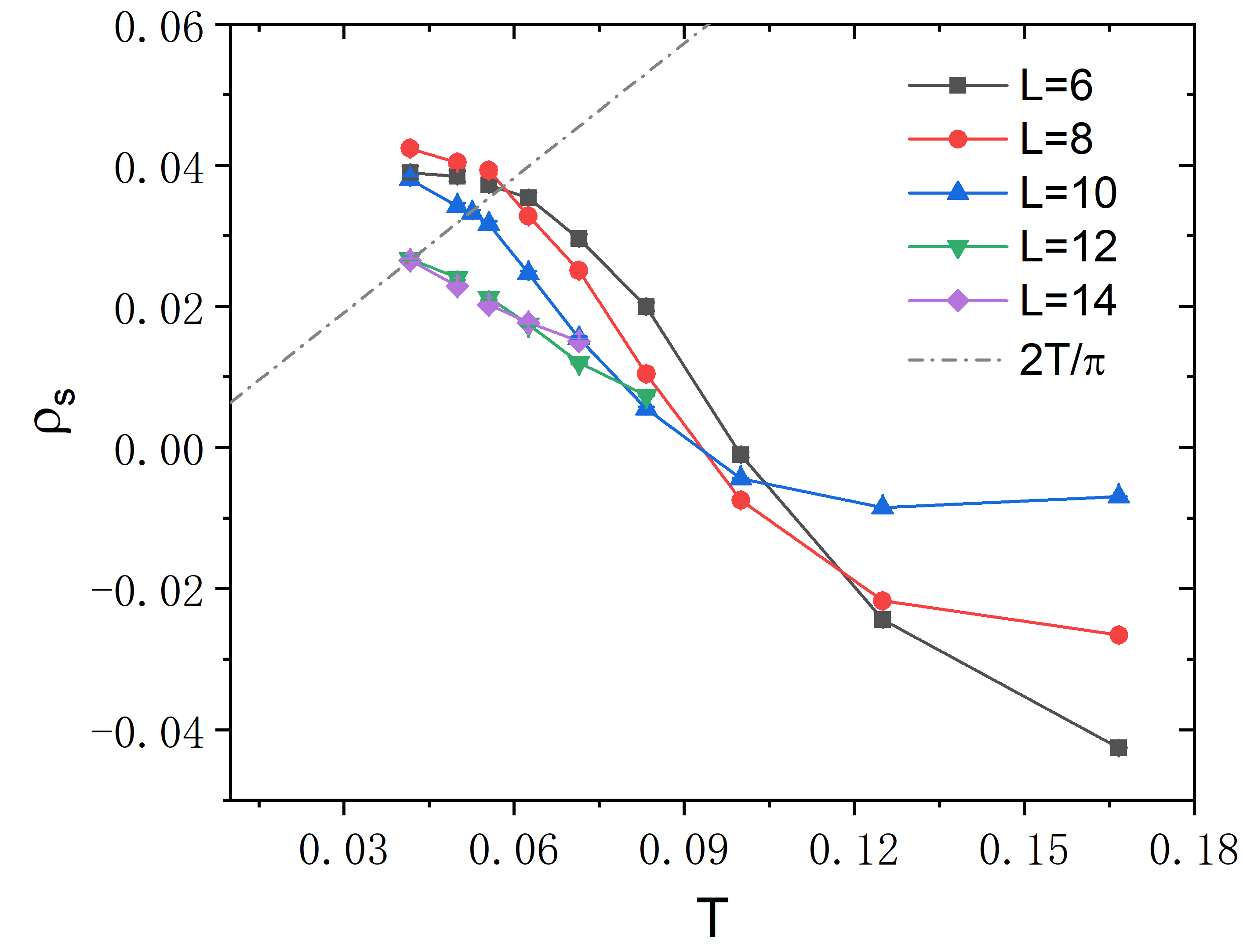}}~~~~	
\subfigure{\label{figS4b}\includegraphics[width= 0.4\linewidth]{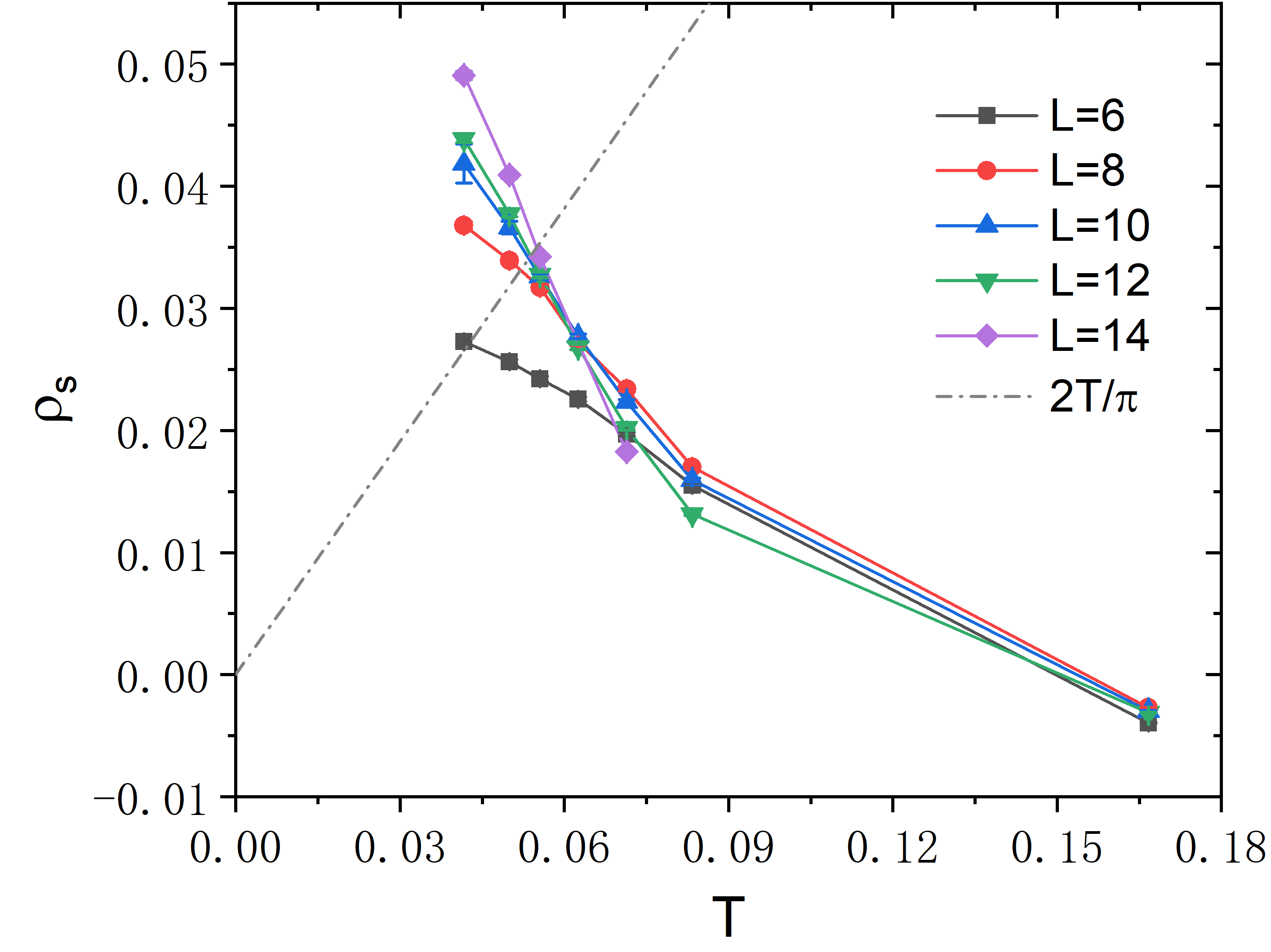}}
\caption{Estimating $\rho_s$ of doped SSH model in the AA limit with weak EPC strength $\lambda=0.2$. (a) QMC results without adding an unit quantum flux. (b) QMC results with the unit flux. From (b) we obtain $T_c=0.042$. }
\label{FigS4}
\end{figure}

\subsection{D. Finite size effect at weak coupling }
For weakly coupled SSH model in AA limit \Eq{EqDefAA2}, the superconducting gap $\Delta$ is small, such that the coherence length $\xi\sim 1/\Delta$ diverges and saturates in small finite lattice, yielding severe finite size effect. The superfluid stiffness for $\lambda=0.2$ is depicted in \Fig{FigS4}(a). The intersections with line $2T/\pi$ are no longer concentrated for different system size. To solve this, we add an unit quantum flux through the lattice for the weak coupling case. The presence of the perpendicular magnetic field will significantly reduce the finite size effect\cite{Assaad_PRB_Kondo2002}. More specifically, we add Peierls phase factor in the hopping amplitude, and modify the Hamiltonian as
\bea\label{EqDefAA3}
H_\mathrm{AA}=-t\sum_{i,\delta,\s}\inc{c^\dagger_{i,\s}c_{i+\delta,\s}\E{\imth A^\s_{i,\delta}}+\mathrm{h.c.}}-\frac{J}{4}\sum_{i,\delta}\inc{\sum_{\s}c^\dagger_{i,\s}c_{i+\delta,\s}\E{\imth A^\s_{i,\delta}}+\mathrm{h.c.}}^2
\eea
with phase factor following Landau gauge:
\bea\label{EqPhaseFac}
A^\s_{i,\delta}=
\begin{aligned}
    \left\{
    \begin{array}{ll}
      -\frac{2\pi i_y}{L^2}\s   & \mathrm{for}\ \delta=x \\
      \frac{2\pi i_x}{L}\s   & \mathrm{for}\ \delta=y\ \mathrm{and}\ i_y=L \\
      0                   & \mathrm{otherwise}
    \end{array}
    \right.
\end{aligned}
\eea
Note that opposite signs are taken for two spin sectors in order to keep the model free of sign problem. The Peierls phase factor will also be introduced into the single particle contribution of the current operator \Eq{EqAACurrent}, while the pairing term in \Eq{EqAACurrent} is not affected, since the phases in two spin sectors compensate with each other. The results of $\rho_s$ estimated from the Hamiltonian \Eq{EqDefAA3} are depicted in \Fig{FigS4}(b). The finite size effect is remarkably reduced, yielding $T_c=0.042$ for $\lambda=0.2$. This result is already contained in \Fig{Fig2}(b) in the main text.
\end{appendices}

\end{document}